\documentclass{aa}  
\usepackage{graphicx}
\usepackage{txfonts}
\usepackage{natbib}
\newcommand{\kms}{\rm {km\ s$^{-1}$}}
\begin{document}

   \title{The counter-rotating stellar core of NGC 4494}

   \author{L. Coccato
          \inst{1}
          \and
          L. Morelli\inst{2}
          \and
          A. Pizzella\inst{3,4}
          \and
          E. M. Corsini\inst{3,4}
          \and 
          V. Cuomo\inst{5}
          }

   \institute{European Southern Observatory, Karl-Schwarzschild-Strasse 2, D-85748 Garching, Germany.
              \email{lcoccato@eso.org}
         \and
             Instituto de Astronom\'ia y Ciencias Planetarias, Universidad de Atacama, Avenida Copayapu 485, 1530000 Copiap\'o, Atacama, Chile.
         \and   
           Dipartimento di Fisica e Astronomia "G. Galilei", Universit\`a di Padova, vicolo dell’Osservatorio 3, I-35122 Padova, Italy.
         \and  
          INAF – Osservatorio Astronomico di Padova, vicolo dell’Osservatorio 5, I-35122 Padova, Italy.
          \and
          Departamento de Astronom\'ia, Universidad de La Serena, Av. Ra\'ul Bitr\'an 1305, La Serena, Chile.
             }

   \date{Received XXX; accepted YYY}

  \abstract
   {Kinematically decoupled cores (KDCs) are often found in the centers of early-type galaxies.}
   { We aim to investigate the kinematics, structure, and stellar populations of the KDC residing in the early-type galaxy NGC 4494 to understand its formation.}
   {We used long-slit spectroscopic data obtained with the FORS2 instrument on the VLT to measure the stellar kinematics and stellar populations. We performed a spectroscopic decomposition to disentangle the properties of the KDC from those of the host galaxy
   and construct models of the observed rotation curve.}
   {The rotation curve is characterized by two symmetric dips at $|R|=6$\arcsec, where the rotation velocity drops to zero. Contrary to previous studies that explained the decoupled structure as a rapidly co-rotating disk, our analysis clearly shows that it is a counter-rotating component. A counter-rotating core is indeed needed to reproduce the observed dip in the velocity curve. The properties of the stellar populations of the decoupled core and the main galaxy are very similar: old stars (12--13 Gyr) with slightly super-solar metallicities (0<[Z/H]<0.15 dex) and $\alpha$-enhanced (0 <[$\alpha$/Fe]< 0.15 dex).}
  {Our results indicate that the counter-rotating component is a disk of about 1 kpc in diameter that is obscured by dust in the central 0.12 kpc. The properties of its stellar populations suggest that it formed from the same material as the main stellar body of the host galaxy. This could have happened via internal processes such as the precession of a pre-existing rotating core, or, alternatively, via gas accretion in retrograde orbits followed by star formation. In the latter scenario, the accretion event occurred almost simultaneously with the formation of the galaxy, using material that had the same composition as the gas from which the stars in the main body of the galaxy were formed.}
  

   \maketitle
  
%

\section{Introduction}
\label{sec:introduction}

Kinematically decoupled cores (KDCs) in galaxies are intriguing stellar structures that often provide valuable insights into the complex formation and evolutionary processes of galaxies. These cores are characterized by distinct kinematic properties from the surrounding galaxy, and they have been known since the 1980s (e.g., \citealt{Efstathiou+82, Bender+88}), but their census and characterization was boosted in the last decades thanks to the advent of integral field spectroscopy (e.g., \citealt{Krajnovic+11}). We refer the reader to 
comprehensive reviews on the subject of kinematic decoupling in galaxies from both observational (\citealt{BertolaCorsini99}, \citealt{Corsini14}) and simulation (\citealt{Khoperskov+21}) points of view.

The first property of a KDC that stands out is the decoupling of its angular momentum with respect to that of the host galaxy. KDCs can co-rotate (as in NGC 4026; \citealt{Krajnovic+11}), counter-rotate (as in NGC 448; \citealt{Nedelchev+19}), or rotate around an orthogonal axis (as in NGC 4698; \citealt{Bertola+99}) with respect to the main stellar body.
Counter-rotating cores, and more generally counter-rotating stellar structures, are typically associated with binary mergers (e.g., \citealt{Crocker+09, Bois+11}) or the accretion of gas onto retrograde orbits that turned into stars (e.g., \citealt{Coccato+13}, \citealt{Bassett+17}, \citealt{Bao+22}). As a consequence of gas accretion, the majority of stellar counter-rotating structures are younger and less massive than the rest of the galaxy with few exceptions (e.g., \citealt{Bevacqua+22}). In most cases, the counter-rotating component shows different metallicity and $\alpha$ enhancement than the main galaxy component (\citealt{Coccato+11, Coccato+13, Pizzella+14, Pizzella+18, Nedelchev+19}; however, see also \citealt{Coccato+15, Morelli+17} for exceptions). Alternative internal scenarios involving the dissolution of a bar \citep{Evans+94} have also been proposed to explain cases where the luminosity and properties of the counter-rotating component are the same of those of the host galaxy.

In this paper, we revisit the case of NGC 4494, a well-studied early-type galaxy in the Coma I group of galaxies, which is known to host a stellar kinematically decoupled core in the inner kiloparsec \citep{Bender+88}. 
Despite its classification as E1--2 elliptical \citep{rc3}, its photometric profile is better fit with the combination of a S\'ersic bulge and an exponential disk about four times brighter \citep{Bonfini+18}. Moreover, it displays the disk-like kinematics typical of fast rotators \citep{Emsellem+11}. Its total mass at virial radius ($r_{\rm vir}=261 \pm 30$ kpc) is $M_{\rm vir} = 1.12^{+0.33}_{-0.24} \cdot 10^{12}$ M$_\odot$, with a dark-matter fraction of $f_{\rm DM} = 0.44 \pm 0.08$ at five effective radii ($\sim 18$ kpc; see \citealt{Napolitano+09}). Its total apparent luminosity in the $K$-band is $K_T=7.00$ \citep{2mass}, which corresponds to an absolute magnitude of $M_K=-24.14$ mag, assuming a distance of 16.9 Mpc (see below).
NGC 4494 has a nuclear dust ring of semi-major axis 1\farcs5 (0.12 kpc), which is interpreted as a signature of an accretion event \citep{vandokkum+95, Forbes+95}. An accretion event, or an interaction in general, could also be responsible for the large-scale kinematic misalignment observed at $\approx 2'$ ($\approx 10$ kpc) in the planetary nebulae population \citep{Coccato+09, Pulsoni+19}.
On the other hand, the  globular cluster population does not display the same misalignment as the planetary nebulae, but it shows an extremely regular velocity field \citep{Brodie+14, Foster+16}.

Regarding the KDC in the core of NGC 4494, previous long-slit spectroscopic analyses, such as that of \citet{Bender+88}, have suggested that it could be attributed to a disk structure that is aligned and co-rotates with the main body of the galaxy, but rotates at significantly higher velocities. This interpretation was also accepted by subsequent integral-field spectroscopic surveys: \citet{Krajnovic+11} describes NGC 4494 as "regular rotation" with "two aligned velocity maxima".
In this paper, we challenge this interpretation, providing evidence for the existence of a counter-rotating stellar core, whose properties and formation mechanism are the subjects of this study.

This paper is organized as follows. In Sect. \ref{sec:sec:observations}, we present the observations and data reduction. In Sect. \ref{sec:rotation_curve}, we describe the data analysis, including the measurements of the rotation curve and the spectral decomposition. We also discuss and compare the two kinematic models for the KDC: co-rotating and counter-rotating stellar structures.
In Sect. \ref{sec:discussion}, we derive the properties of the KDC and discuss its formation scenario. In Sect. \ref{sec:summary}, we summarize our findings.
In the paper we adopt a distance to NGC 4494 of 16.9 Mpc
using a Hubble constant of $H_0 = 74.4$ km s$^{-1}$ Mpc$^{-1}$ \citep{Tully+13}, which leads to a spatial scale of 0.082 kpc arcsec$^{-1}$.

\section{Observations and data reduction}
\label{sec:sec:observations}
The long-slit spectroscopic data of NGC 4494 were acquired with the FOcal Reducer/low dispersion Spectrograph 2 (FORS2, \citealt{Appenzeller+}) mounted on the Unit Telescope 1 (Antu) of the Very Large Telescope (VLT) at the European Southern Observatory (ESO) in Paranal, Chile, under the program 76.B-0788A (P.I., L. Coccato).
FORS2 was equipped with the 1400V grism and the 0\farcs5 slit, providing a  spectral resolution
at full width at half maximum (FWHM) of FWHM $\simeq 135$ \kms\ at 5400 \AA\ (as measured on the reduced arc reference frames), and a wavelength coverage of 4650--5900 \AA.
The $2\times2$ pixel binning ensured a dispersion sampling of 0.64 \AA\ pixel$^{-1}$ and a spatial sampling along the slit of $0\farcs25$ pixel$^{-1}$. 

Ten exposures of 1375 seconds each along the galaxy major axis (position angle PA=$0^\circ$) were taken for a total time on target of $\simeq 3.8$ hrs. Data were acquired in service mode under dark time conditions with a median seeing of FWHM $\simeq$ 1\arcsec,\, as measured with the differential image motion monitor. The advantages of the FORS2 dataset with respect to other long-slit or integral-field observations available in the literature are the higher spectral and spatial resolution that allow us to constrain the properties of the kinematically decoupled core.
The observation results were originally published in \citet{Coccato+09}, with the aim of linking the kinematics in different regions of the galaxy obtained with different tracers. In this work, we repeat the data reduction using  an updated version of the FORS2 pipeline and a different cosmic-ray detection tool. We also used a set of stellar libraries optimized for the study of the stellar populations, which was not done in the 2009 work.
The data reduction (up to the sky subtraction) is performed with the FORS2 pipeline version 5.5.3 run under the EsoReflex environment \citep{Freudling+13}. The spectra taken with the two detectors are combined considering the gap between them. The cosmic rays are detected on the individual spectra using the ESO implementation\footnote{\url{https://ftp.eso.org/pub/dfs/pipelines/instruments/hdrldemo/}, version 1.4.0.} of the  Laplacian cosmic-ray identification algorithm \citep{vanDokkum01}. The spectra are aligned using the bright galaxy continuum as reference and stacked, masking the contribution of bad pixels and cosmic rays.
The sky background contribution is evaluated at the edges of the detector where the contribution of the galaxy was negligible by fitting a first-order polynomial along the slit direction and then subtracted. The fitting procedure includes an iterative 3$\sigma$ clipping rejection to remove outliers from the linear fit along the slit direction.

The comparison with the previous reduction does not show substantial differences, except for a slightly lower velocity dispersion in the outer regions\footnote{The difference is at most 5 \kms. It arises from the different approximation adopted for the instrumental FWHM: 140 \kms\ in \citet{Coccato+09} and 135 \kms\ in the current analysis. However, such a difference has no impact on the analysis nor the results of this work.} and fewer bad pixels in the final stacked spectrum due to the improved FORS2 pipeline and cosmic-ray detection. We therefore decided to use the new reduction for further analysis, although the measured stellar line-of-sight distribution is consistent with the previous work within
errors.

\section{Analysis of the stellar kinematics}
\label{sec:rotation_curve}
The spectra are logarithmically binned along the dispersion direction with a velocity scale of 35 \kms\ pixel$^{-1}$ and then binned along the long-slit direction in order to ensure a minimum signal-to-noise ratio\footnote{The signal-to-noise ratio was measured with the Python implementation of {\tt DER\_SNR},  
\url{https://www.stecf.org/software/ASTROsoft/DER_SNR/}.} of 150 per pixel in the fitting spectral range (4750 -- 5870 \AA). 
In order to extract the stellar kinematics, the binned spectra are  fit with the penalized pixel fitting method ({\tt pPXF}, \citealt{Cappellari+04}) using a set of stellar templates from the MILES library \citep{Vazdekis+10}, which accounts for enhancement of $\alpha$ elements and has a spectral resolution of FWHM=2.5 \AA\ \citep{Falcon+11}. We performed the fit in several ways, as detailed in the next sections, according to the different kinematic models we explored.  

No emission lines are detected in the studied wavelength region, despite NGC 4494 being known to host a rapidly rotating \ion{H}{I} disk ($\Delta V = 552$ km s$^{-1}$, \citealt{Garcia+94}), and it has an H$\alpha$ disk that extends out to 1 kpc along the photometric major axis (i.e., same size as the KDC), but with unknown kinematics \citep{Hermosa+22}.
On the other hand, the presence of CO emission is uncertain; \citet{Sage+07} claimed a broad and weak detection in CO(1--0), but this is not visible in the CO(2--1) spectrum, whereas \citet{Young+11} did not find any signature of CO emission and claimed a non-detection.

\begin{figure*}
    \centering
    \includegraphics[width=0.95\textwidth]{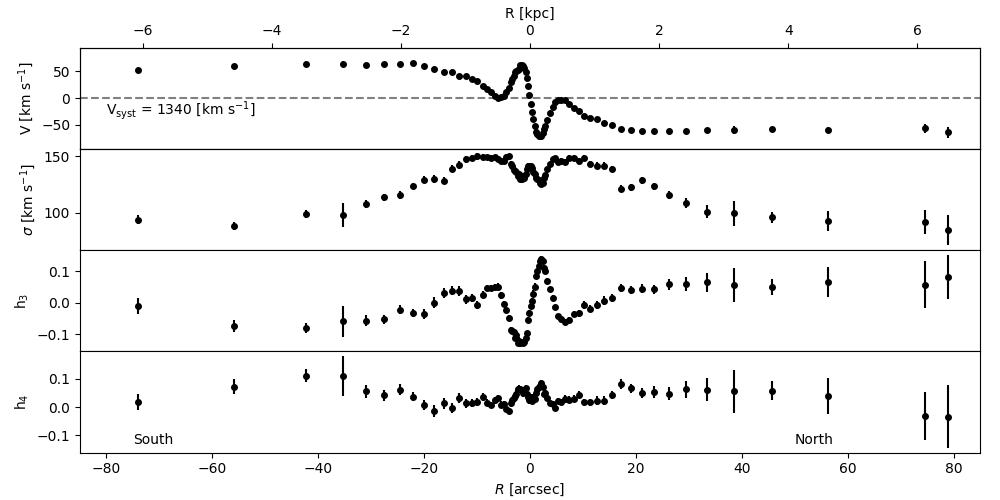}
    \caption{Line-of-sight stellar kinematics of NGC 4494 along $\rm PA=0^\circ$. From top to bottom: Radial profiles of the velocity, velocity dispersion, and Gauss-Hermite moments $h_3$ and $h_4$. The systemic velocity $V_{\rm syst}$ is computed as the median of the measured velocities. The orientation of the long-slit is given in the bottom panel.}
    \label{fig:single_component_fit}
\end{figure*}

\subsection{The "single-component" fit}
\label{sec:stellar_kinematics}

We parameterized the line-of-sight stellar velocity distribution (LOSVD) with a Gaussian function plus $h_3$ and $h_4$ Gauss-Hermite coefficients. We refer to this fit as the "single-component" fit. The spectra are fit in the 4750 -- 5870 \AA\ wavelength region using a first-order additive polynomial degree (to compensate for additive bias in sky subtraction) and a fourth-order multiplicative polynomial degree (to compensate for inaccuracies of the response curve). 
The results do not vary significantly when changing the degree of the polynomials. 
The radial profiles along the slit of the stellar velocity, velocity dispersion, $h_3$ and $h_4$ coefficients are shown in Fig. \ref{fig:single_component_fit}. 

As evident from Fig. \ref{fig:single_component_fit}, the kinematics in the inner $|R| \lesssim$ 20\arcsec\ are indicative of the presence of a kinematically decoupled stellar component. The rotation curve steeply rises, reaching the maximum amplitude $V_{\rm max} \simeq 70$ km s$^{-1}$ at $|R|=$1\farcs5. It then decreases back to the systemic velocity at $|R|=6$\arcsec, it rises back until reaching $V_{\rm max}$  at $|R|\simeq 20$\arcsec\ again, and then it slowly declines, probably due to a slight misalignment between the orientation of the long-slit (PA$=0^\circ$) and kinematic major axes (PA$_{\rm kin}=185^\circ$, \citealt{Krajnovic+11}). The stellar-velocity dispersion exhibits two peaks that are symmetric with respect to the center, corresponding to the dip in the rotation curve at 6", and two local minima with slightly different values corresponding to the peaks in the rotation curve.

\subsection{The "two-component" fit}

In this section, we assume the presence of two distinct kinematic components in NGC 4494: the KDC and the main galaxy body. Thus, we performed a two-component fit with {\tt pPXF}. We considered and compareed two models for the KDC.
In the first model (Sect. \ref{sec:cr_model}), we postulate the presence of a stellar disk, characterized by a steep velocity gradient, that is co-rotating with the main body of the host galaxy. This is the model that has been assumed so far for NGC 4494 (e.g., \citealt{Bender+88, Krajnovic+11}).
In the second model (Sect. \ref{sec:ring_model}), we postulate the presence of a stellar component that is counter-rotating with respect to the host galaxy and that is confined between $1\farcs5<|R|<20\arcsec$. This implies that this counter-rotating component has zero flux within $|R|<1\farcs5$, such as a ring or a torus. 

In both scenarios, we used a "toy model" to  describe the observed rotation curve. Then, we used the LOSVD derived from the toy model as our initial predictions in a fitting procedure aimed at finding the properties of the stellar components. The toy model is constructed by the combination of two arctan-like rotation curves:

\begin{equation}
    V_i(R) = \frac{2a_0}{\pi} \arctan\left( R/a_1 \right) \cdot \left( a_2 R^2/|R|+1 \right) + a_3
    \label{eqn:toy}
,\end{equation}
 with $i=1,2$. $V_1$ corresponding to the rotation curve of the host galaxy and $V_2$ to that of the KDC (either a fast co-rotating disk or a counter-rotating component).
The first part of Eq. \ref{eqn:toy} describes an "arctan-type" rotation curve with amplitude $a_0$ and  scale radius $a_1$. The second part  describes a linearly decreasing trend of the rotation curve with gradient $a_2$, and accounts for the misalignment between the orientations of the slit and the kinematic major axis. Finally, $a_3$ is the systemic velocity. The choice of the analytic form is purely empirical: it is only used to provide a good description of the data. The best-fitting model is used as starting guess in the spectral decomposition (Sects. \ref{sec:cr_model} and \ref{sec:ring_model}). In principle, any other analytic expression could be used, as long as it reproduces the observed rotation curve.
Each component has its flux profile $F_1(R)$ and $F_2(R)$. The ratio $f_{21}(R)$ of the flux profiles is determined so that the final observed rotation curve $V(R)$ is given by their flux-weighted average:

\begin{eqnarray}
    V(R) &=& \frac{V_1(R)F_1(R) + V_2(R)F_2(R)}{F_1(R)+F_2(R)}  \nonumber\\
         &=& \frac{V_1(R) + V_2(R)f_{21}(R)}{1+f_{21}(R)}
         \label{eqn:ratio}.
\end{eqnarray}

\subsubsection{The co-rotating stellar-disk model}
\label{sec:cr_model}
\begin{figure}
    \includegraphics[width=0.45\textwidth]{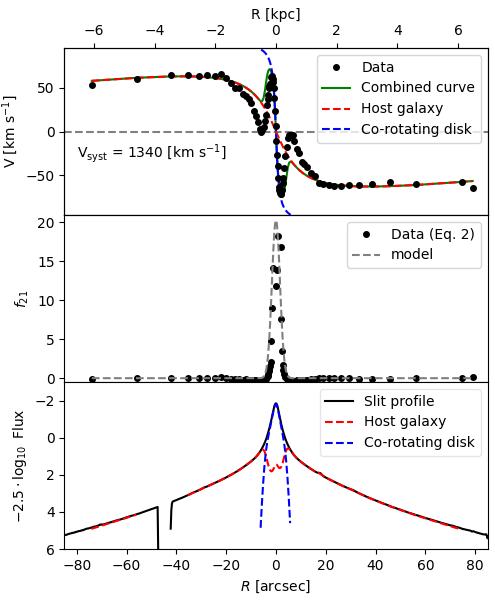}
    \caption{Toy model of a co-rotating stellar disk. 
    Top panel: Measured radial profiles of line-of-sight velocity (black circle) and best-fitting rotation curves for the co-rotating disk (dashed blue line) and the host galaxy (solid red line) as determined by Eq. \ref{eqn:toy}. Middle panel: Flux ratio $f_{12}$ as determined from Eq. \ref{eqn:ratio} (black circle) and its best-fitting Gaussian model (gray dashed line). Bottom panel: Surface-brightness profiles of the galaxy along the slit (black solid line, obtained by integrating the observed spectrum along the wavelength direction), of the KDC (dashed blue line) and the host galaxy (dashed red line) as obtained assuming the best-fit model for their flux ratios. The radial profile of $f_{12}$ in the middle panel indicates that the co-rotating stellar disk dominates the total flux within 6\arcsec.}
    \label{fig:cr_model_toy}
\end{figure}

As stated in Sect. \ref{sec:introduction}, this model has been the main interpretation of the observed kinematics of NGC 4494. We call this model the co-rotating disk model. Figure \ref{fig:cr_model_toy} shows the toy model where the super-imposition of the rotation curves of the two components (in green) partially reproduces the measured rotation curve.

The toy model is determined by fitting Eq. \ref{eqn:toy} for each component in the radial range in which it dominates. The best-fitting rotation curves are shown in the top panel of Fig. \ref{fig:cr_model_toy}.
Then, using Eq. \ref{eqn:ratio}, the flux ratio $f_{21}$ is determined, which is then fit by one Gaussian function. The empirical values of $f_{21}$ and its best-fitting model are shown in the middle panel of Fig. \ref{fig:cr_model_toy}. The combined rotation curve obtained by the superposition of the two components, with the flux ratio given by the best-fitting Gaussian model of $f_{21}$ is shown in the top panel of Fig. \ref{fig:cr_model_toy}. The toy model is not able to reproduce the dip of the rotation curve, unless we assume nonphysical negative values of the flux distribution for one of the two components.

\begin{figure}
    \includegraphics[width=0.45\textwidth]{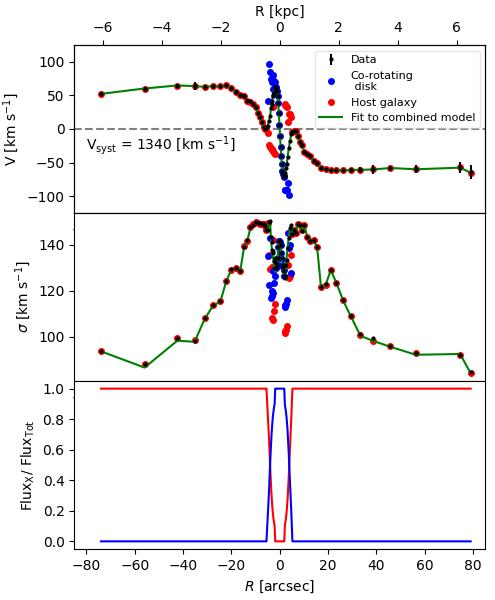}
    \caption{
     Best-fitting model of a co-rotating stellar disk. From top to bottom: Radial profiles of line-of-sight velocity, velocity dispersion, and flux ratio of each component with respect to total. The black circles correspond to data, blue circles to the co-rotating disk, red circles to the host galaxy, and the solid green line to the single-component fit of the best-fitting model obtained by combining the best-fitting spectra of the two components. Blue and red lines shows the fraction of total flux for the co-rotating disk and host galaxy, respectively. The best fit unveils the presence of counter-rotating stars, despite the starting guesses derived from the best-fitting toy model representing two co-rotating systems.}
    \label{fig:best_fit_disk_model}
\end{figure}

As reference, we also show in the bottom panel of Fig. \ref{fig:cr_model_toy} the surface-brightness profiles of the two components, assuming the best fit for $f_{21}(R)$ and the surface-brightness profile of the galaxy as observed through the FORS2 long slit. Appendix \ref{sec:coefficients} reports the best-fitting coefficients of the rotation curves  (Table \ref{tab:toy}) and the function used to model the flux ratio $f_{12}$.

The LOSVD obtained from the best-fitting parameters of the toy model is used as starting guess to fit the observed spectra. We set up {\tt pPXF} to fit each observed spectrum with two distinct kinematic stellar components. Both components share the same library of templates. 
Given the small difference between the kinematics of the two components, there is a high degree of degeneracy between the fitting parameters. Therefore, we are forced to constrain the range of some variables.
The starting guesses for the velocity of the two components are taken from the toy model, allowing $\pm 50$ \kms\ of freedom to cover the gap between the models (blue and red lines in Fig. \ref{fig:cr_model_toy}) and the measurements (black circles in Fig. \ref{fig:cr_model_toy}). The flux ratio between the two different stellar kinematic components is kept fixed,\footnote{We also tested the case where we leave the flux ratio free to vary. In this case, the fit routine tends to go back to the single component fit. It gives most of the weight to one component, whose rotation curve follows the one measured in Fig. \ref{fig:single_component_fit}.} as determined in the toy model of Fig. \ref{fig:cr_model_toy}, and it describes the case where the main body of the galaxy hosts a much brighter co-rotating disk in its center.
The fit results are shown in Fig. \ref{fig:best_fit_disk_model}.
As a self-consistency check, we fit the best-fitting model (obtained as superposition of the two best-fitting stellar components) with a single component as in Sect. \ref{sec:stellar_kinematics}. This is represented by the green line in Fig. \ref{fig:best_fit_disk_model}.

In order to reproduce the observations, one of the best-fitting stellar components must counter-rotate with respect to the other, despite the starting guesses describing two co-rotating components.
In conclusion, it is not possible to reproduce the observations, in particular the shape of the stellar velocity curve, with a model that includes two  co-rotating stellar components.

\subsubsection{The counter-rotating stellar component model}
\label{sec:ring_model}
\begin{figure}
    \centering
    \includegraphics[width=0.45\textwidth]{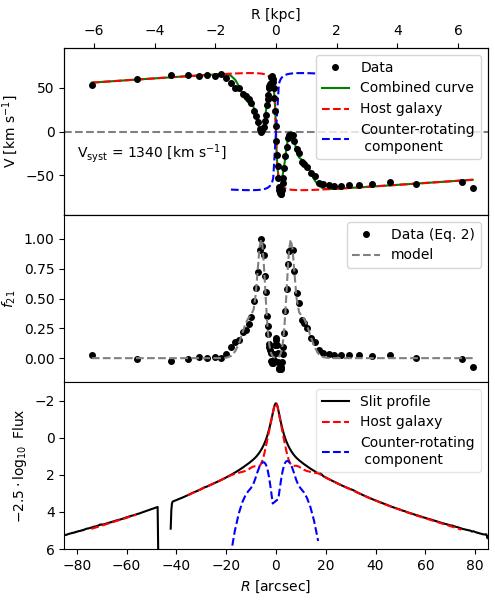}
    \caption{As in Fig \ref{fig:cr_model_toy}, but for a counter-rotating stellar component. The radial profile of $f_{12}$ in the middle panel indicates that the contribution of the counter-rotating component is null within the central $|R| <1\farcs5 $ and then rises and matches the flux of the host galaxy at $|R|\simeq 6$\arcsec. Outwards, it declines until disappearing for $|R|>18$\arcsec.}
    \label{fig:ring_model_toy}
\end{figure}

This model represents a new interpretation of the kinematic properties of the core of NGC 4494. Contrary to the previous works, we postulate the presence of a counter-rotating stellar component, as luminous as the main stellar body and confined between $1\farcs5$<$|R|$<$20$\arcsec, that is the region where the rotation velocity drops to zero and the velocity dispersion profile displays two peaks. 
This model implies that the counter-rotating component has zero flux in the inner $1\farcs5$ (as in the case of a ring or a centrally obscured disk); therefore, we refer to this model as the counter-rotating ring model. We discuss the morphology of the secondary component in more detail in Sect. \ref{sec:discussion}. Figure \ref{fig:ring_model_toy} shows a toy model where the superposition of the rotation curves of the two components (in green) nicely reproduces the observations.

\begin{figure}
    \centering
    \includegraphics[width=0.45\textwidth]{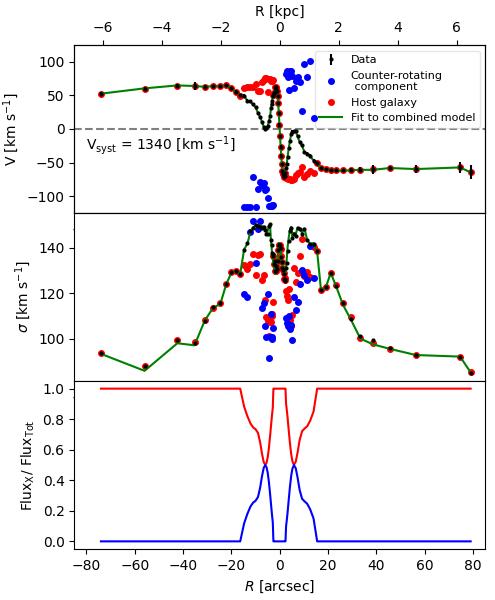}
    \caption{As in Fig. \ref{fig:best_fit_disk_model}, but for a counter-rotating stellar ring.}
    \label{fig:best_fit_ring_model}
\end{figure}

In the toy model, the rotation curve of the counter-rotating component is assumed to be symmetric to the one of the host galaxy.
The toy model is determined by fitting Eq. \ref{eqn:toy} for each component in the radial range in which it dominates. The best-fitting rotation curves are shown in the top panel of Fig. \ref{fig:ring_model_toy}.
Then, using Eq. \ref{eqn:ratio}, we determine the flux ratio $f_{21}$, which we then fit with two Gaussian functions, both symmetric with respect to $R=0$. The empirical values of $f_{21}$ and its best-fitting model are shown in the middle panel of Fig. \ref{fig:ring_model_toy}. The combined rotation curve obtained by the superposition of the two components, with the flux ratio given by the best-fitting double-Gaussian model of $f_{21,}$ is shown in the top panel of Fig. \ref{fig:ring_model_toy}. Unlike the previous co-rotating model, the counter-rotating model is capable to reproduce the dip in the rotation curve.
For reference, we also show the surfac- brightness profiles of the two components in the bottom panel of Fig. \ref{fig:ring_model_toy}, assuming the best fit for $f_{21}$ and the surface-brightness profile of the galaxy as observed through the long slit.

The LOSVD obtained from the best-fitting parameters of the toy model is used as an initial estimation to fit the observed spectra.
We set up {\tt pPXF} to fit each observed spectrum with two distinct kinematic stellar components. Both components share the same library of stellar templates. 
As in the previous model, there is a high degree of degeneracy between the different fitting parameters. Therefore, we are forced to constrain the range of some variables.
The velocity of the main galaxy is kept constrained within $\pm 10$ \kms\ of the guess of the toy model; the velocity of the counter-rotating component is allowed to vary from the toy model within $\pm 50$ \kms. The flux ratio between the two different stellar kinematic components is kept fixed as determined by the double-Gaussian fit of the toy model. The fit results are shown in Fig. \ref{fig:best_fit_ring_model}.
As a self-consistency check, we fit the best-fitting model (obtained as superposition of the two best-fitting stellar components) with a single component as in Sect. \ref{sec:stellar_kinematics}. This is represented by the green line in Fig. \ref{fig:best_fit_ring_model}.

The rotation curve of the KDC is very noisy, but it is possible to recognize some general trends. The rotation amplitude is larger than that of the main galaxy body, and there is an indication of decreasing rotational velocity for $|R|>10$\arcsec. 
In conclusion, the existence of a centrally obscured 
counter-rotating stellar component is necessary to explain the observations.

\section{The properties and the origin of the KDC}
\label{sec:discussion}

\subsection{Stellar populations}
\label{sec:stellar_population}
\begin{figure}
    \centering
    \includegraphics[width=0.45\textwidth]{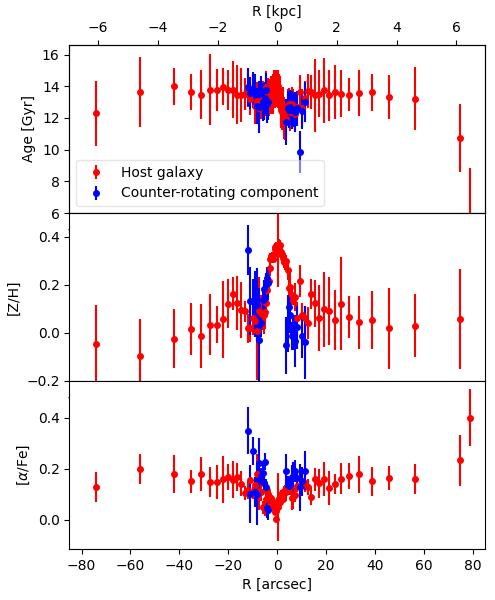}
    \caption{Radial profile of mass-weighted stellar age (top panel), metallicity (middle panel), and $\alpha$ enhancement (bottom panel) of the counter-rotating stellar ring (blue circles) and host galaxy (red circles).}
    \label{fig:ssp}
\end{figure}

\begin{figure*}
    \centering
    \includegraphics[width=0.95\textwidth]{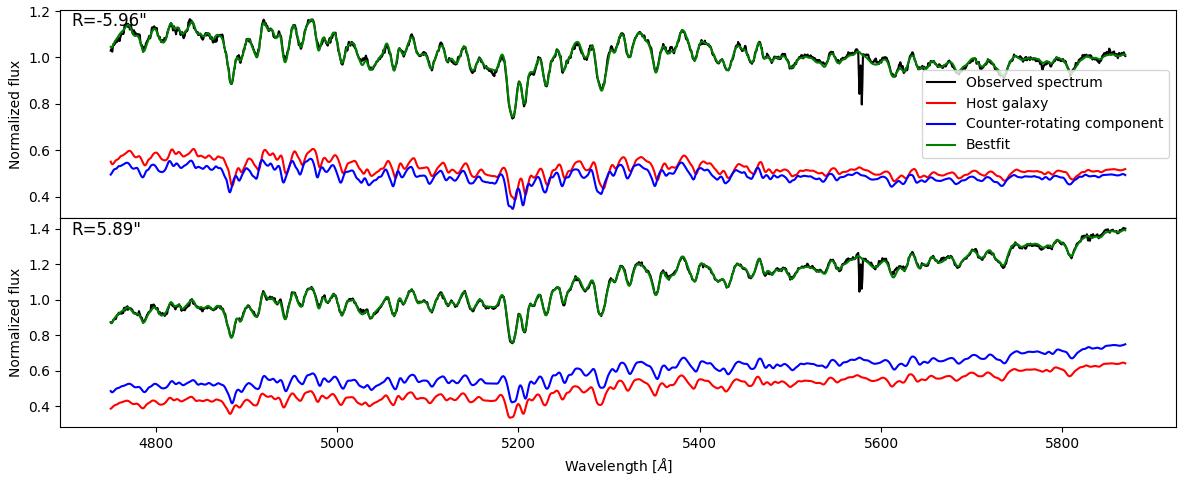}
    \caption{Spectral decomposition of galaxy spectrum at $R=-5\farcs96$ (upper panel) and $+5\farcs89$ (lower panel), where the contribution of the counter-rotating component is maximum. The observed spectrum (in the observer-frame, black line) is modeled by the sum (green line) of the contributions of two stellar components, corresponding to the counter-rotating stellar ring (blue line) and host galaxy (red line). The observed spectrum is normalized at its median value. No emission lines are visible in the galaxy spectrum across the entire slit. The artifact at 5577 \AA\ is due to over-subtraction of the \ion{O}{I} sky emission line at 5577 \AA.}
    \label{fig:fit_example}
\end{figure*}

Figure \ref{fig:ssp} shows the radial trends of the mass-weighted values of age, metallicity, and $\alpha$ enhancement of the two stellar components as obtained from the {\tt pPXF} best-fitting model. At each radius of each component, the properties of the best-fitting templates are averaged using the corresponding weights determined by {\tt pPXF}. The errors are computed by means of Monte Carlo simulations.
Overall, the properties of the stellar populations of the host galaxy and the counter-rotating component are consistent within the uncertainties. This is also seen in Fig. \ref{fig:fit_example}, where we show that the spectra of the two stellar components, as obtained from the spectral decomposition, indeed look very similar.
Both components are very old (12--13 Gyrs) and their metallicity slightly super-solar (0<[Z/H]<0.15 dex, where they co-exist). The metallicity of the main component at the center, where the secondary component is not detected, increases to up to 0.4 dex. Their $\alpha$ enhancement ($0<$[$\alpha$/Fe]$<0.15$ dex) is consistent with a star formation timescale that ranges between 1 and 4 Gyr (using Eq. 4 from \citealt{Thomas+05}). 
Our results indicate that the two components must have either originated from the same stellar population with a mechanisms that decoupled their angular momentum or formed simultaneously from two counter-rotating gas clouds with the same chemical composition. 

\subsection{Morphology: Possibility of a genuine ring or a centrally obscured ring}
\label{sec:morphology}
The small velocity separation between the kinematically decoupled core and the host galaxy generates a high degeneracy between the model parameters. Therefore, it is not possible to have a fully unconstrained fit of our data, and we had to constrain some parameters. This limits our freedom in interpreting the results, because some best-fitting parameters depend on our assumptions. However, it is clear that in order to explain the fact that the rotation curve goes back to the systemic velocity at about $|R| \simeq 6$\arcsec\ from the center, a model including two counter-rotating components is needed. The exact shape and amount of velocity of the two components within $|R|<15$\arcsec\ slightly depends on the model assumptions. However, in all the explored cases, the contribution of the decoupled core to the total galaxy light has the following overall behaviour: it is null in the centre, it is maximum about $|R| \pm 6$\arcsec\ (i.e., where the dips in the rotation curve and the peaks in the velocity dispersion profile are measured), and then it goes down again. The presence of the central hole in the flux distribution of the secondary component leads us to interpret it as a ring. However, it is not possible to infer the two-dimensional morphology of the counter-rotating stellar component using only long-slit data. They actually only reveal the presence of two peaks in the flux distribution along the slit, which are symmetric with respect to the galaxy center. The transition from "two peaks" to "ring" is supported by the two-dimensional kinematic map presented in \citet{Krajnovic+11}, which shows the presence of an azimuthally symmetric component. In order to investigate the two-dimensional morphology of the counter-rotating structure, we modeled the HST-WFC2 images taken with filters F814W and F555W that are available in the  \textit{Hubble} Legacy Archive (Appendix \ref{sec:imaging}).
We fit two S\'ersic profiles, representing the bulge and the disk of the galaxy, and analyzed the residuals. In the fit, we masked the foreground stars, the dust ring, and the region dominated by the KDC, with the goal of having its image popping up in the residuals. 
The residual images reveal the presence of complex structures in the central $|R| \lesssim 8$\arcsec, but nothing that looks like a ring or any other axisymmetric structure. In particular, the region at $|R| \simeq 6$\arcsec, where the KDC should be as bright as the host galaxy, no excess of light in the residual images is detected (Fig. \ref{fig:galfit}).

Despite there being no direct evidence of the presence of a stellar ring in the HST images of NGC 4494 nor in the residuals of the photometric decomposition, the "ring" scenario naturally explains the properties of the rotation curve. Stellar rings are common among galaxies of all morphological types, with a large variety of sizes (see \citealt{Buta+96} for a review). Albeit rare, counter-rotating stellar rings have already been observed (e.g., \citealt{Silchenko+06, Martinsson+18}). However, stellar rings are typically associated with bars (e.g., \citealt{Fernandez+21}), ionized gas, and recent star formation episodes (e.g., \citealt{Sarzi+07}), and they are bluer than the underlying stellar population (e.g., \citealt{Buta+96, Buta+15}). However, these  properties are not observed in NGC 4494. 

A possible alternative to the ring scenario is the presence of a centrally obscured disk. As stated in Sect. \ref{sec:introduction}, NGC 4494 hosts a dust ring (e.g., Fig. \ref{fig:galfit}) whose semi-major axis of $\sim 1\farcs5$ coincides with the "hole" in the light distribution of the counter-rotating structure. In this case, the counter-rotating disk must have similar vertical height to that of the dust and much smaller than that of the host galaxy; otherwise, the dust ring would equally attenuate the light from the galaxy. This condition seems to be fulfilled, because the stellar rotation is larger and the velocity dispersion is smaller in the counter-rotating component than in the host galaxy (Fig. \ref{fig:best_fit_ring_model}). This indicates that the KDC is more rotationally supported and, thus, more likely to be thinner than the main galaxy component.

\subsection{Origin of the stellar counter-rotating component}

The observed long-slit kinematics of NGC 4494 is better explained with the presence of a counter-rotating stellar structure (either a ring or a centrally obscured disk) than a fast, co-rotating stellar disk. We may ask ourselves how and when such a component formed.
The mean age of the stars in the counter-rotating component is about 12 Gyr, which is coeval with that of the main body of the galaxy. This sets the formation epoch of the counter-rotating stars, if not the assembly epoch of the component itself, to at least 12 Gyr ago with a star-formation timescale that ranges between 1 and 4 Gyr. In addition, any formation mechanism must also be consistent with the fact that the two stellar components have the same metallicity.
We discuss two possibilities here: an external origin or an internal origin for the counter-rotating component in NGC 4494.

\subsubsection{External origin}

As discussed in Sect. \ref{sec:introduction}, counter-rotating stellar structures are typically associated with accretion events.
In the specific case of NGC 4494, accretion events have been invoked to explain the presence of the nuclear dust ring that is observed in the innermost $1\farcs5$, which is located exactly inside the hole of the counter-rotating structure, be it a ring or an obscured disk. However, dust rings live at most 2-3 Gyr (e.g., \citealt{Comeron+10} and references therein), whereas our analysis reveals that the stars in the counter-rotating component are much older and must have formed almost simultaneously to the stars of the galaxy's main body (see Sect. \ref{sec:stellar_population}).
Therefore, if the dust ring is the signature of an accretion event, it must be independent of the KDC formation and must not be associated with a subsequent major merger or strong interaction, which would have destroyed the pre-existing KDC (\citealt{Bois+11, Sarzi+16}; however, see also \citealt{Ebrova+21} on the survival of KDCs after major mergers).

If the KCD were a counter-rotating ring, then we would have to assume that NGC 4494 hosted a nuclear bar. Indeed, the formation of stellar rings is typically associated to bars. The bar can dissolve, whereas the rings survive \citep{Athanassoula96}. In barred galaxies, gas can accumulate around resonances and then form stellar rings, which can survive after the gas reservoir is depleted. On the other hand, the presence of a past bar is not required if the KDC is a counter-rotating disk.
Whether the counter-rotating component is a ring or an obscured disk, the accreted material must have had the same properties as the gas that formed the stars of the galaxy main body; in addition, the star formation should have occurred several gigayears ago, in an epoch very close to the formation of the stars of the galaxy's main body, as the age difference between the two components is not evident in the galaxy spectra.

Galaxy mergers with mass ratios of 2:1 and 1:1 can also generate elliptical galaxies with a stellar counter-rotating KDC \citep{Bois+11}, but the merger remnants are slow rotators, whereas NGC 4494 is classified as fast rotator. In addition, the nuclear kinematic feature observable in the velocity fields of merger remnants of \citet{Bois+11} are just the tip of the iceberg of a much larger structure that extends as far as the host galaxy.
These counter-rotating structures are typically less luminous than the main galaxy component (except for the inner regions, where the KDC dominates the galaxy luminosity), and thus the overall kinematics is driven by the contribution of the host galaxy. If their surface brightness is at least $\approx$20\% of that of the host galaxy, they can be identified by means of spectroscopic decomposition (as in, e.g., \citealt{Coccato+13}, \citealt{Nedelchev+19}), but this seems not to be the case of NGC 4494. Finally, in the binary merger scenario, the two progenitors must have had stars with the same age, metallicity, and $\alpha$ enhancement to explain the observed properties of NGC 4494.

\subsubsection{Internal origin}

The fact that the stars of the KDC have the same properties as those of the host galaxy (at least in the region where the two components overlap) suggests that they originated from the same parent stellar population. Thus, an internal process could explain the origin of the KDC. Moreover, internal scenarios favor the formation of a counter-rotating disk, as they do not involve the presence of gas trapped into bar resonances to form the ring.

In their simulations, \citet{Ebrova+21} showed that it is possible to create a KDC from the precession of a previously existing  rotating core. The core could be formed by the very same gas that generated the galaxy, without the need to invoke external gas accretion onto counter-rotating orbits. The precession can be triggered by a pericentric passage or by the flyby of another galaxy, or without an obvious external cause. Depending on when precession started and on its velocity, the KDC can be counter-rotating with respect to the host galaxy.

Alternatively, \citet{Evans+94} proposed the dissolution of a bar as formation scenario for those cases where the stellar populations and fraction of retrograde stars are equal to those of the host galaxy. After the dissolution of the bar, the stars moving into box orbits escape from their confining potential well and move onto tube orbits. Since there are as many clockwise box orbits as counter-clockwise box orbits, the process generates two counter-rotating stellar structures with the same luminosity and stellar population properties (see also  \citealt{Athanassoula96}). 
Unfortunately, while this scenario explains our findings well, and the fast rotator kinematics and surface-brightness profile of NGC 4494 might suggest a morphological transformation toward an earlier type, there is no evidence to support the hypothesis that NGC 4494 was barred, nor are we aware of any simulations that study the creation of counter-rotating components induced by bar dissolution in detail.

\section{Summary}
\label{sec:summary}
In this paper, we analyze the stellar kinematics along the major axis of NGC 4494. We demonstrate that the peculiar shape of the rotation curve (a steep central rise followed by a decline of rotation down to zero) is better explained by a counter-rotating component rather than a fast, co-rotating decoupled inner disk as assumed by previous works. This result opens the question as to whether other galaxies, previously thought to host a fast co-rotating stellar core, could actually host a counter-rotating  component instead.

Although the exact morphology (ring or centrally obscured disk) and the formation mechanism (external or internal) of the counter-rotating component in NGC 4494 cannot be unambiguously determined, our analysis indicates that it is more likely a centrally obscured disk than a ring.
 In order to match the observed kinematics, the disk must be obscured in the central $|R|< 1\farcs5$. This condition is fulfilled by the presence of a dust ring of that size. Also, the vertical height of the disk must be smaller than that of the main disk, otherwise the dust would also obscure the galaxy's main body. This condition is consistent by the smaller velocity dispersion and larger rotational velocity of the decoupled component with respect to the host galaxy.

The age, metallicity, and $\alpha$ enhancement of the counter-rotating disk are the same as in the main body of the galaxy. This suggests that the counter-rotating disk either originated from an internal process, such as the precession of a pre-existing rotating core, or from an early accretion of gas into retrograde orbits followed by star formation.
In the latter scenario, the formation of counter-rotating stars has to be coeval to the formation of the stars in the galaxy main body and has to come from gas with exactly the same chemical composition of the one that gave birth to the stars in the host galaxy.
A caveat to our interpretations is given by the fact that we do not observe any clear evidence of a stellar disk or ring by analyzing the available images of NGC 4494.

The knowledge of the kinematics of the gas component (either ionized or neutral) would put more constraints on the origin of the KDC in NGC 4494, under the assumption that the gas component was not accreted after the formation of the counter-rotating component. Indeed, if the gas is found to rotate in the same direction as the main stellar body, it would further support the internal origin scenario because the precession affects only the stellar component. On the other hand, if the gas is found to counter-rotate with respect to the main stellar body, it would further support the external accretion scenario.

\begin{acknowledgements}
Based on observations collected at the European Southern Observatory under ESO programme 76.B-0788(A).
This research used the HLA facilities of the STScI, the ST-ECF and the CADC with the support of the following granting agencies: NASA/NSF, ESA, NRC, CSA.
EMC and AP are supported by the Istituto Nazionale di Astrofisica (INAF) through the grant Progetto di Ricerca di Interesse Nazionale (PRIN) 2022 C53D23000850006 and Padua University with the grants Dotazione Ordinaria Ricerca (DOR) 2020-2022.
VC acknowledges the support provided by Chilean ANID through 2022 FONDECYT postdoctoral research grant no. 3220206.

\end{acknowledgements}

\bibliographystyle{aa}
\bibliography{NGC4494} 

\begin{appendix}
    \section{Coefficients of the toy model stellar components}
    \label{sec:coefficients}
 In this section we list the best-fitting parameters of the analytic functions that describe the rotation curves and flux distributions of the two stellar components in the toy models. These analytic expressions are used as starting guesses in the spectral decomposition fit.
Parameters of rotation curves are give in Table \ref{tab:toy}.
The analytic expression of the flux ratio $f_{12}$ (dashed grey lines in Figs. \ref{fig:cr_model_toy} and \ref{fig:ring_model_toy}) are:

\begin{itemize}
    \item {Co-rotating disk model}. The flux ratio is modelled as a single Gaussian function:
    \begin{equation}
        f_{12}(R) = a_0 \exp^{-0.5\left( R/a_1\right)^2} \nonumber
        \end{equation}
        with $a_0=20.2$6 and $a_1=1.62$\arcsec.
        \bigskip
    \item {Counter-rotating component model}. The flux ratio is modelled by two Gaussian functions, each of them symmetric with respect to the slit centre $R=0$\arcsec:
    \begin{equation}
 f_{12}(R) = a_0 \exp^{-0.5\left( \frac{|R|-a_1}{a_2}\right)^2} + b_0 \exp^{-0.5\left( \frac{|R|-b_1}{b_2}\right)^2}\nonumber
 \end{equation}
    with $a_0=0.863$, $a_1=5.813$\arcsec, $a_2=1.494$\arcsec, $b_0=0.345$, $b_1=10.30$\arcsec, and $b_2=3.197$\arcsec.
\end{itemize}
 
 \begin{table}[h]
      \caption[]{Best-fitting parameters of toy models rotation curves defined by Eq. \ref{eqn:toy}.}
       \label{tab:toy}
         \begin{tabular}{lcc}
            \hline
            \noalign{\smallskip}
            Coefficient      &  {\rm Host\ galaxy} & {\rm Decoupled component} \\
            \noalign{\smallskip}
            \hline \\
\noalign{\centering{Co-rotating disk model}      }
            \noalign{\smallskip}
            $a_0$ [\kms]      & 83.7 & 104.3 \\
            $a_1$ [arcsec]    & 8.17 & 0.93 \\
            $a_2$ [\kms arcsec$^{-1}$]       & --0.00353 & {\tt --} \\
            $a_3$ [\kms] & 1340 & 1339 \\
            \noalign{\smallskip}\\
\noalign{\centering{Counter-rotating component model}}
            \noalign{\smallskip}
            $a_0$ [\kms] & 70.6 & -70.6 \\
            $a_1$ [arcsec]    & 0.46 & 0.46 \\
            $a_2$ [\kms arcsec$^{-1}$]  & -0.0028 & -0.0028 \\
            $a_3$ [\kms] & 1340 & 1340 \\
            \noalign{\smallskip}
            \hline
         \end{tabular}
   \end{table}

\section{Imaging}
\label{sec:imaging}

\begin{figure*}
    \centering
    \includegraphics[width=0.99\textwidth]{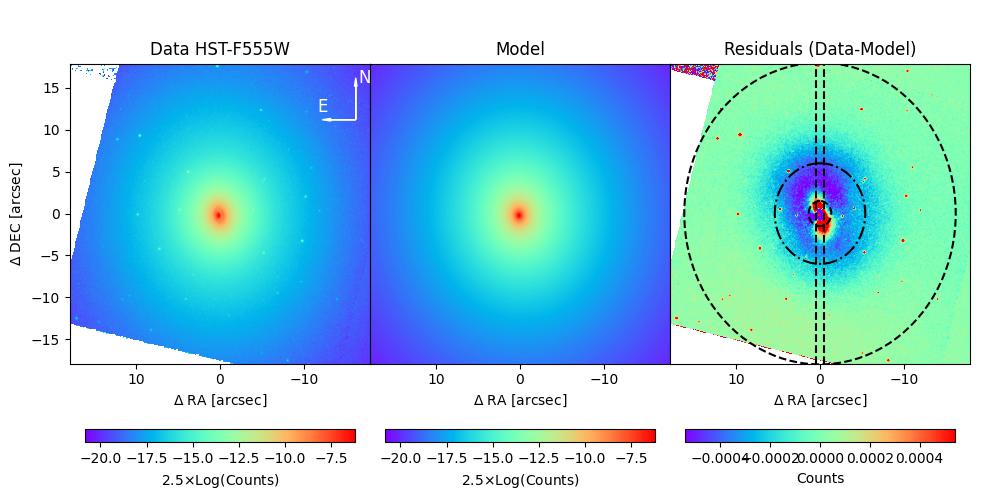}
    \bigskip
    \includegraphics[width=0.99\textwidth]{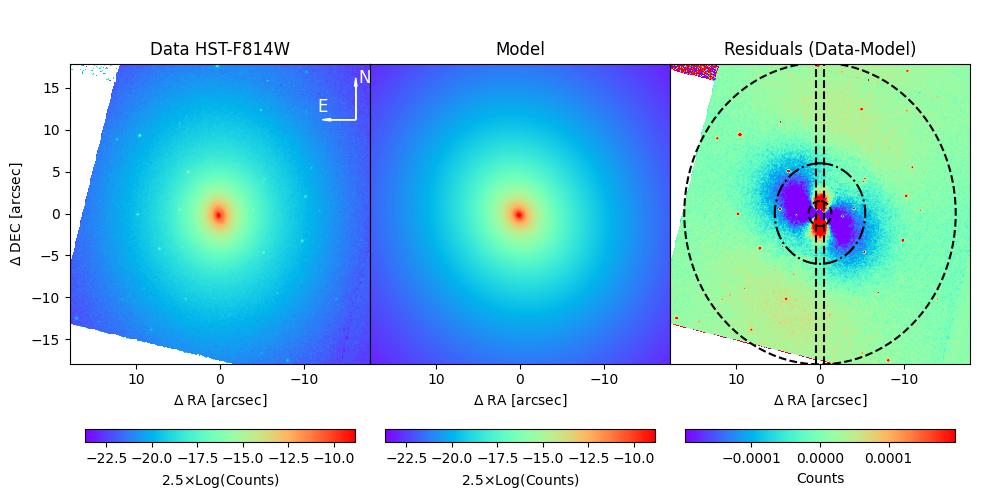}
    \caption{Zoomed-in view of the inner regions of the HST image of NGC 4494 (left panels), its best-fitting model by {\tt galfit} (central panels), and residuals (right panels). The top and bottom panels show the observations obtained with filters F555W and F814W, respectively. The arrows indicate the orientation on the sky. The scale is 0.082 kpc arcsec$^{-1}$. Dashed ellipses marks the location where the counter-rotating component should be located: inner ellipse has $1\farcs5$ semi-major axis (where the signal of the counter-rotating component becomes relevant), central ellipses has 6\arcsec\ semi-major axis (where the flux contribution of the counter-rotating component is maximum), outer ellipses has 18\arcsec\ semi-major axis (where the signal of the counter-rotating component is negligible). Dashed vertical lines mark the position of the FORS2 long-slit.}
    \label{fig:galfit}
\end{figure*}

We inspect the archival images from the \textit{Hubble} Space Telescope obtained with the WFPC2 camera mounting filters F555W and F814W at spatial resolution of $0\farcs1$ pixel$^{-1}$. The fully processed and combined images are directly retrieved from the \textit{Hubble} Legacy Archive facility.
We model the light distribution with the superposition of two S\'ersic profiles and a tilted plane (to include residual additive contribution from data reduction or contamination from nearby sources) with the {\tt galfit} version 3.0 photometric decomposition code \citet{Peng+10}. Foreground stars are modelled with a two-dimensional Gaussian function (with independent dispersions and orientations in the spatial directions) and used as the point spread function in the {\tt galfit} fit. The original image, the best fit model and the residuals
are shown in Fig. \ref{fig:galfit}. 
We repeat the exercise also for FORS2 acquisition images and for \textit{Spitzer} (IRAC camera 4.5 $\mu$m and 3.6, $\mu$m at spatial resolution of $0\farcs75$ pixel$^{-1}$), but no evident disk- or ring like structures are observed. Both FORS2 acquisition images and IRAC images are not shown in this paper.

\end{appendix}

\end{document}